\documentstyle[aps,prl,floats,epsfig]{revtex}  
\draft 

\begin{document}


\title{\Large Observation of $B^+ \to \chi_{c0} K^+ $}
\author{The Belle Collaboration \\
\begin{center}
\tighten
  K.~Abe$^{8}$,               
  K.~Abe$^{37}$,              
  R.~Abe$^{27}$,              
  I.~Adachi$^{8}$,            
  Byoung~Sup~Ahn$^{15}$,      
  H.~Aihara$^{39}$,           
  M.~Akatsu$^{20}$,           
  Y.~Asano$^{44}$,            
  T.~Aso$^{43}$,              
  V.~Aulchenko$^{2}$
% ...
\end{center}
\small
\begin{center}
% ...
$^{2}${Budker Institute of Nuclear Physics, Novosibirsk}\\
% ...
$^{8}${High Energy Accelerator Research Organization (KEK), Tsukuba}\\
% ...
$^{15}${Korea University, Seoul}\\
% ...
$^{20}${Nagoya University, Nagoya}\\
% ...
$^{27}${Niigata University, Niigata}\\
% ...
$^{37}${Tohoku Gakuin University, Tagajo}\\
% ...
$^{39}${University of Tokyo, Tokyo}\\
% ...
$^{43}${Toyama National College of Maritime Technology, Toyama}\\
$^{44}${University of Tsukuba, Tsukuba}\\
% ...
\end{center}
}

\maketitle
\normalsize
\tighten

\begin{abstract}

  Using a sample of $31.3\times10^{6}$ $B\bar{B}$ pairs collected
with the Belle detector at the $\Upsilon$(4S) resonance, we make the 
first observation of the charged $B$ meson decay to $\chi_{c0}$ and
a charged kaon. The measured branching fraction is
${\cal{B}}(B^+\to \chi_{c0}K^+)=(6.0^{+2.1}_{-1.8}\pm1.1)\times10^{-4}$,
where the first error is  statistical, and the second is systematic.\\

\end{abstract}

\pacs{PACS numbers: 13.25.Hw, 14.40.Nd }

{\renewcommand{\thefootnote}{\fnsymbol{footnote}}}
\setcounter{footnote}{0}

\twocolumn
\normalsize


   Two-body decays of $B$ mesons with a charmonium particle in the 
final state have recently received substantial attention due to their 
sensitivity to CP violation in $B$ system. The production rate of 
charmonium states in $b\to c\bar{c}s$ transitions also provides valuable
insight into the dynamics of strong interactions in heavy meson systems.
For instance, although the production of the $\chi_{c0}$ $P$-wave $0^{++}$ 
state in $B$ decays vanishes in the factorization approximation as a 
consequence of spin-parity and vector current conservation,
it is possible if there is an exchange of an additional 
soft gluon~\cite{beneke,diehl}. At present, only upper limits on the 
$B\to\chi_{c0}K$ branching fractions exist~\cite{edwards}. In this 
Letter, we report the first observation of the $B^+\to\chi_{c0}K^+$ 
decay mode. The analysis is performed using data collected with the 
Belle detector at the KEKB asymmetric energy $e^+e^-$ collider~\cite{KEKB}.
The data sample consists of 29.1~fb$^{-1}$ taken at the $\Upsilon(4S)$
resonance containing $31.3\times10^{6}$ $B\bar{B}$ pairs, 
and 2.3~fb$^{-1}$ taken 60~MeV below the $\Upsilon(4S)$ resonance to 
perform systematic studies of the $e^+e^-\to q\bar{q}$ background.


    The Belle detector~\cite{NIM} is a large-solid-angle magnetic 
spectrometer that consists of a three-layer silicon vertex 
detector (SVD), a 50-layer central drift chamber (CDC) for charged 
particle tracking and specific ionization measurement ($dE/dx$), 
an array of aerogel threshold \v{C}erenkov counters (ACC), 
time-of-flight scintillation counters (TOF), and an array of 8736 
CsI(Tl) crystals for electromagnetic calorimetry (ECL) located inside 
a superconducting solenoid coil that provides a 1.5~T magnetic field. 
An iron flux return located outside the coil is instrumented to 
detect $K_L$ mesons and to identify muons (KLM). 
Electron identification is based on a combination of CDC $dE/dx$
measurements, the response of the ACC, and the position, shape and 
energy deposition of the associated ECL shower. 
We use a Monte Carlo (MC) simulation to model the response of the 
detector and determine acceptance~\cite{GEANT}.


  The analysis of the $B^+\to\chi_{c0}K^+$ decay is performed in the 
framework of a general study of the $B^+\to K^+h^+h^-$ decay, where 
$h$ stands for either pion or kaon~\cite{khh} (charge conjugation is 
implied throughout this Letter). We reconstruct the $\chi_{c0}$ meson 
in the decay modes $\chi_{c0}\to\pi^+\pi^-$ and $\chi_{c0}\to K^+K^-$.
Charged tracks are required to satisfy requirements based on the 
average hit residual and on their impact parameters relative to the 
interaction point. We require that the transverse momentum of the track 
be greater than 0.1 GeV/$c$ to reduce low momentum combinatorial 
background. For charged hadron identification, we use a combination of 
CDC $dE/dx$ measurements, flight time measured in TOF, and the response 
of the ACC. We select kaon candidate tracks with a set of criteria that 
has about 90\% efficiency, a charged pion misidentification probability 
of about 8\%, and a negligible contamination from protons. 
We reject tracks that are positively identified as electrons.

   We reconstruct $B$ mesons by combining a $\chi_{c0}$ with a charged 
kaon. The candidate events are identified by their center-of-mass (c.m.) 
energy difference, \mbox{$\Delta E=(\sum_iE_i)-E_b$}, and the beam 
constrained mass, $M_{bc}=\sqrt{E_b^2-(\sum_i\vec{p}_i)^2}$, where 
$E_b = \sqrt{s}/2$ is the beam energy in the c.m. frame, and $\vec{p}_i$ 
and $E_i$ are the c.m. three-momenta and energies of the candidate $B$ 
meson decay products. We select events with $M_{bc}>5.20$~GeV/$c^2$
and $|\Delta E|<0.2$~GeV, and define a {\it signal} region of
$|M_{bc}-M_B|<9$~MeV/$c^2$ and $|\Delta E|<0.04$~GeV and two $\Delta E$
{\it sideband} regions $-0.08$~GeV~$<\Delta E<-0.05$~GeV and 
$0.05$~GeV~$<\Delta E<0.15$~GeV~\cite{dEreg}.


   To suppress the large combinatorial background, which is dominated 
by the two-jet-like $e^+e^-\to~q\bar{q}$ continuum process, we use 
variables that characterize the event topology. We require 
$|\cos\theta_{\rm thr}|<0.80$, where $\theta_{\rm thr}$ is the angle 
between the thrust axis of the $B$ candidate and that of the rest of 
the event. This eliminates 83\% of the continuum background and retains 
79\% of the signal events. We also define a Fisher discriminant, 
${\cal F}$, that includes the production angle of the $B$ candidate, 
the angle of the $B$ candidate thrust axis with respect to the beam 
axis, and nine parameters that characterize the momentum flow in the 
event relative to the $B$ candidate thrust axis in the c.m. 
frame~\cite{VCal}. We impose a requirement on ${\cal{F}}$ that rejects 
79\% of the remaining continuum background and retains 74\% of the signal.
In the case where the $\chi_{c0}$ is reconstructed in the $K^+K^-$ mode,
the continuum background is much smaller and a looser requirement that 
rejects 53\% of the continuum background with about 89\% efficiency 
for the signal is used.

   As shown in Ref.~\cite{khh}, the background to the $K^+\pi^+\pi^-$ 
final state from $B\bar{B}$ decays is dominated by decays of the type 
$B^+\to [K^+\pi^-]\pi^+$, where $[K^+\pi^-]$ denotes an intermediate state 
that can decay into the $K^+\pi^-$ such as $K^{*0}(892)$ or $\bar{D^0}$.
To suppress this type of background, we require that the invariant mass of 
the $K^+\pi^-$ system be greater than 2.0~GeV/$c^2$. For the $K^+K^+K^-$ 
final state, we require that the invariant mass for both $K^+K^-$ 
combinations  be greater than 2.0~GeV/$c^2$ to suppress the background 
from charmless $B$ decays.

\begin{figure}[t]
  \centering
  \includegraphics[width=8.6cm]{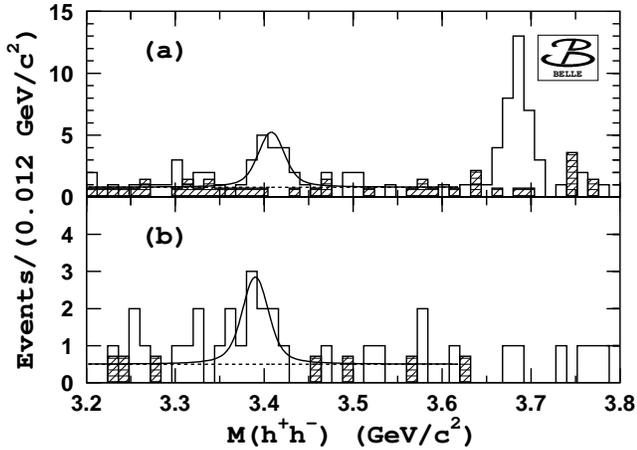}
  \centering
  \caption{The (a) $\pi^+\pi^-$ and (b) $K^+K^-$ invariant mass spectra. 
           Open histograms correspond to events from the $B$ signal region,
           and hatched histograms correspond to events from the 
           $\Delta E$ sidebands. The curves are fits to the signal data.}
  \label{hhmass}
\end{figure}

   We select all $K^+\pi^+\pi^-$ ($K^+K^+K^-$) combinations from the 
$B$ signal region that satisfy the selection criteria described above 
and have a $\pi^+\pi^-$ ($K^+K^-$) invariant mass in the range
$3.2$~GeV/$c^2$~$<M(h^+h^-)<3.8$~GeV/$c^2$. The resulting $\pi^+\pi^-$ 
and $K^+K^-$ invariant mass spectra are shown in Fig.~\ref{hhmass}.
Since in the case of the $K^+K^+K^-$ final state there are two kaons 
with the same charge, we distinguish the $K^+K^-$ combinations with the 
smaller and larger invariant masses. Here, we only plot the larger of 
the two possible combinations and require the invariant mass of the 
smaller combination to be less than 3.35~GeV/$c^2$. The peaks near 
3.4~GeV/$c^2$ in Figs.~\ref{hhmass}(a) and~\ref{hhmass}(b) are 
identified as the $\chi_{c0}$ meson. The peak at 3.69~GeV/$c^2$ in 
Fig.~\ref{hhmass}(a) corresponds to the $\psi(2S)$ meson from the decay 
mode $B^+\to \psi(2S)K^+$, $\psi(2S)\to \mu^+\mu^-$ with muons 
misidentified as pions. The hatched histograms in Fig.~\ref{hhmass} 
correspond to the events from the $\Delta E$ sidebands plotted with a 
weight of 0.62.

   These $\pi^+\pi^-$ and $K^+K^-$ spectra are fitted in the range 
3.20~GeV/$c^2$~$<M(h^+h^-)<3.62$~GeV/$c^2$ to the sum of a constant for 
the background and a Breit-Wigner function convolved with a Gaussian 
resolution function for the signal. The width of the resolution function 
is fixed at $11.0$ MeV/$c^2$ as determined from a fit to the $J/\psi$ 
peak in the $\mu^+\mu^-$ invariant mass spectrum. The full width of the 
Breit-Wigner function is fixed at the world average $\chi_{c0}$ width of 
14.9~MeV~\cite{PDG}. The results of the fit are given in Table~\ref{hhfit}.
The statistical significance, $\Sigma$, of the signal, in terms of the 
number of standard deviations, is calculated as 
$\sqrt{-2\ln({\cal{L}}_0/{\cal{L}}_{\rm max})}$,
where ${\cal{L}}_{\rm max}$ and ${\cal{L}}_{0}$ denote the maximum likelihood 
with the nominal signal yield and with the signal yield fixed at zero, 
respectively.

  The peak position in the $K^+K^-$ spectrum is found to be shifted below 
the PDG value~\cite{PDG} by $25\pm10$~MeV/$c^2$: see Fig.~\ref{hhmass}(b) 
and Table~\ref{hhfit}. Although this shift is consistent with a statistical 
fluctuation, we note that, as shown below, we observe evidence for a 
non-resonant-like signal in the $B^+\to K^+K^+K^-$ final state. As a result, 
the $K^+K^-$ invariant mass distribution in the $\chi_{c0}$ region could be 
distorted by the effects of interference with an amplitude not related to 
the $B^+\to\chi_{c0}K^+$. Because of this uncertainty, we base our branching 
fraction measurement on the $\chi_{c0}\to \pi^+\pi^-$ decay mode only.

   For the branching fraction calculation, we normalize our results 
to the observed  $B^+\to J/\psi K^+$, $J/\psi\to \mu^+\mu^-$ signal. 
This removes systematic effects in the particle identification efficiency,
charged track reconstruction efficiency and the systematic uncertainty 
due to the cuts on event shape variables. To avoid additional systematic 
uncertainty in the muon identification efficiency, we do not use muon 
identification information for $J/\psi$ reconstruction. Instead, we apply 
the same pion-kaon separation requirement for muons from the $J/\psi$ as 
for pions from the $\chi_{c0}$. The feed-across from the $J/\psi\to~e^+e^-$ 
submode is found to be negligible (less than 0.5\%) after the application 
of the electron veto requirement.

\begin{table}[t]
\caption{Results of the fit to the $\pi^+\pi^-$, $K^+K^-$ and $\mu^+\mu^-$
         invariant mass spectra. }
\medskip
\label{hhfit}
  \begin{tabular}{lcccc}
                                    & Eff.
                                    & Peak
                                    & Fit yield 
                                    & $\Sigma$ \\
      Channel  & (\%) & (GeV/$c^2$) & (events) & ($\sigma$)  \\ \hline
  $\chi_{c0}\to\pi^+\pi^-$ & $21.0$ & $3.408\pm0.006$ 
                                    & $16.5^{+5.6}_{-4.8}$ & $4.4$ \\
  $\chi_{c0}\to K^+K^-$    & $12.9$ & $3.390\pm0.010$ 
                                    & $ 8.7^{+4.3}_{-3.6}$ & $3.0$ \\ \hline
  $J/\psi\to\mu^+\mu^-$    & $26.5$ & $3.096\pm0.001$ 
                                    & $406\pm21$           & $ - $ \\
  \end{tabular}
\end{table}

   To determine the $B^+\to \chi_{c0}K^+$ branching fraction, we use the 
signal yield obtained from the fit to the $\pi^+\pi^-$ invariant mass spectrum.
The number of $J/\psi K^+$ signal events is determined from the fit to the 
$\mu^+\mu^-$ invariant mass spectrum (see Table~\ref{hhfit}). Combining all 
the relevant numbers from Table~\ref{hhfit} and using the intermediate 
branching fractions of
${\cal{B}}(\chi_{c0}\to\pi^+\pi^-)=(5.0\pm0.7)\times10^{-3}$ and 
${\cal{B}}(J/\psi\to\mu^+\mu^-)=(5.88\pm0.10)\times10^{-2}$~\cite{PDG}, 
we find the ratio of branching fractions:
\[ \frac{{\cal{B}}(B^+\to \chi_{c0}K^+)}{{\cal{B}}(B^+\to J/\psi K^+)}=
 0.60^{+0.21}_{-0.18}\pm0.05\pm0.08,\]
where the first error is statistical, the second is systematic, and the third 
is due to the uncertainty in the $\chi_{c0}\to\pi^+\pi^-$ branching fraction. 
Here the statistical error includes the errors on the number of signal 
$\chi_{c0}K^+$ and  $J/\psi K^+$ events. The systematic error consists of 
the uncertainty in the $J/\psi\to \mu^+\mu^-$ branching fraction (1.7\%) and
the uncertainty in the background parameterization in the fit to the 
$\pi^+\pi^-$ spectra (7.8\%). Using the world average value of 
${\cal{B}}(B^+\to J/\psi K^+)=(10.0\pm1.0)\times 10^{-4}$~\cite{PDG},
we translate our measurement into the branching fraction:
 \[{\cal{B}}(B^+\to \chi_{c0}K^+)=(6.0^{+2.1}_{-1.8}\pm1.1)\times10^{-4},\]
where the first error is statistical, and the second is the total systematic 
error including uncertainties in 
the $\chi_{c0}\to\pi^+\pi^-$ and $B^+\to J/\psi K^+$ branching fractions.

\begin{figure}[t]
  \centering
  \includegraphics[width=8.6cm]{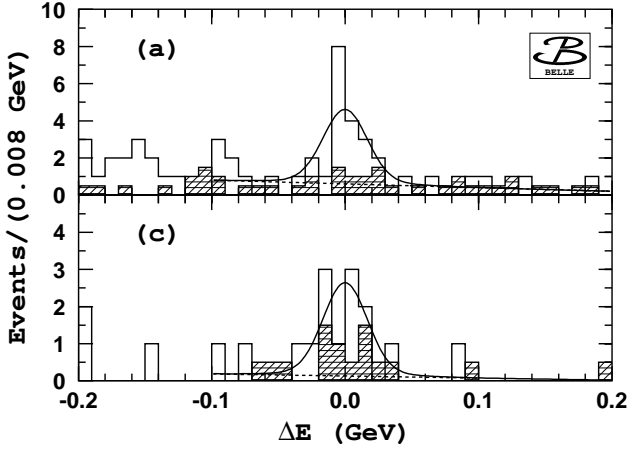}
  \centering
  \caption{The $\Delta E$ distributions for $B^+ \to \chi_{c0}K^+$ candidates;
           (a)-$\chi_{c0}\to\pi^+\pi^-$ and (b)-$\chi_{c0}\to K^+K^-$.
           The hatched histograms correspond to the $\chi_{c0}$ mass sidebands.
           The solid lines display the signal plus background combined shape.
           The dashed lines correspond to the background shape only.}
  \label{dembplot}
\end{figure}

   Figure~\ref{dembplot} shows the projections of the $\Delta E$ signal bands 
for the selected $B^+\to\chi_{c0}K^+$ candidates with 
$|M(h^+h^-)-M_{\chi_{c0}}|<0.05$~GeV/$c^2$ and 
$5.270$~GeV/$c^2$~$<M_{bc}<5.288$~GeV/$c^2$. The hatched histograms in 
Fig.~\ref{dembplot} correspond to the events from the $\chi_{c0}$ mass 
sidebands defined as 
$0.07$~GeV/$c^2$~$<|M(h^+h^-)-M_{\chi_{c0}}|<0.17$~GeV/$c^2$
and plotted with a weight of 0.50. For the $K^+\pi^+\pi^-$ final state, the 
distribution for the sideband events is consistent with background. 
In contrast, in the three charged kaon final state we observe a substantial 
signal for events in the $\chi_{c0}$ mass sidebands. This is the evidence 
for non-resonant-like $B^+\to K^+K^+K^-$ decays that may be responsible for 
the shift in the $\chi_{c0}\to K^+K^-$ mass peak discussed above. A study of 
the full $K^+K^+K^-$ Dalitz plot~\cite{khh} also supports this conclusion.
We represent the background shape in $\Delta E$ with a linear function
and restrict the fit to the range $-0.1$~GeV~$<\Delta E<0.2$~GeV~\cite{dEreg}.
The signal shape is parameterized by a sum of two Gaussians with the 
same mean. The $\Delta E$ shape for the signal is determined from the 
$B^+\to \bar{D^0}\pi^+$ events. The $B^+\to\chi_{c0}K^+$ where 
$\chi_{c0}\to\pi^+\pi^-$ ($K^+K^-$) signal yield of 
$16.4^{+5.2}_{-4.6}$ ($9.7^{+4.0}_{-3.8}$) events obtained from the fit to 
the $\Delta E$ distributions agrees with that obtained from the fit to 
$\pi^+\pi^-$ ($K^+K^-$) invariant mass spectrum.

\begin{figure}[t]
  \includegraphics[width=8.5cm]{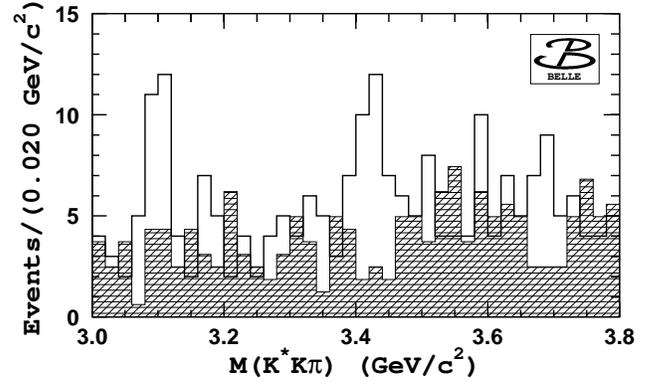}
  \centering
  \caption{The $K^*(892)K\pi$ invariant mass spectrum for 
           events from the $B$ signal region (open histogram) and for events 
           from the $\Delta E$ sidebands (hatched histogram).}
  \label{kkpp}
\end{figure}

   To cross-check the result, we also reconstruct the $\chi_{c0}$ meson 
in the $K^+K^-\pi^+\pi^-$ final state. We reduce the large combinatorial 
background by using events where at least one $K\pi$ pair has an invariant
mass within 75~MeV/$c^2$ of $M_{K^*}$. We also apply the tighter requirement
$|\cos\theta_{\rm thr}|<0.6$ to suppress the continuum background. 
Figure~\ref{kkpp} presents the $K^*(892)K\pi$ invariant mass spectrum 
for the selected events from the $B$ signal region shown by open histogram 
and for events from the $\Delta E$ sidebands shown by hatched histogram 
plotted with a weight of 0.62. For the branching fraction calculation we 
normalize our result to the signal observed in the 
\begin{figure}[t]
  \includegraphics[width=8.5cm]{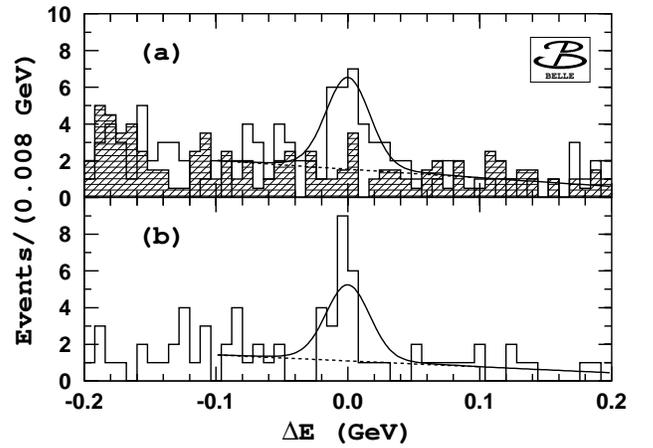}
  \centering
  \caption{The $\Delta E$ distributions for (a) $B^+ \to \chi_{c0}K^+$,
           $\chi_{c0}\to K^{*0}(892)K^-\pi^+$ candidates and (b)
           $B^+ \to J/\psi K^+$, $J/\psi\to K^{*0}(892)K^-\pi^+$ candidates.
	   The hatched histogram in (a) corresponds to events from the
	   $\chi_{c0}$ mass sidebands.
           The solid lines display the signal plus background combined shape.
           The dashed lines correspond to the background shape only.}
  \label{check}
\end{figure}
$B^+\to J/\psi K^+$, 
$J/\psi\to K^{*}(892)K^-\pi^+$ decay mode. Figure~\ref{check} shows the 
$\Delta E$ distributions for $B^+\to \chi_{c0}K^+$ and $B^+\to J/\psi K^+$ 
candidates. From the fit to the $\Delta E$ distributions we find $21.9\pm6.3$
and $18.0\pm5.2$ signal events, respectively. The statistical significance 
of the $B^+\to \chi_{c0}K^+$ signal in this mode is 4.3$\sigma$. Using the 
intermediate branching fractions of
${\cal{B}}(\chi_{c0}\to K^{*0}(892)K^-\pi^+ + c.c.)=
(1.2\pm0.4)\times10^{-2}$~\cite{PDG,Correct} and
${\cal{B}}(J/\psi\to K^{*0}(892)K^-\pi^+ + c.c.)=
(6.3\pm2.1)\times10^{-3}$~\cite{mark},
we determine the ratio of branching fractions:
${\cal{B}}(B^+\to \chi_{c0}K^+)/{\cal{B}}(B^+\to J/\psi K^+)=
0.64\pm0.26\pm0.30$, where the first error is statistical, and the second 
consists of the uncertainty in the secondary branching fractions. 
The obtained number is in agreement with that determined from the 
$\chi_{c0}\to\pi^+\pi^-$ decay mode. We do not include this result in the 
final value for the $B^+\to\chi_{c0}K^+$ branching fraction because of the 
large systematic uncertainty.


  In summary, we report the first observation of the $B^+\to\chi_{c0}K^+$. 
The statistical significance of the signal is 6$\sigma$ when the 
$\chi_{c0}\to\pi^+\pi^-$ and $\chi_{c0}\to K^{*}K\pi$ modes are combined. 
The measured branching fraction is 
${\cal{B}}(B^+\to \chi_{c0}K^+)=(6.0^{+2.1}_{-1.8}\pm1.1)\times10^{-4}$
which is comparable to those for the $B^+\to J/\psi K^+$ and
$B^+\to \chi_{c1}K^+$ decays. This provides evidence for a significant 
nonfactorizable contribution in $B$ to charmonium decay processes.
The result reported here supersedes the previous value based on
21.3~fb$^{-1}$ as reported in Ref.~\cite{chi0}.


   We wish to thank the KEKB accelerator group for the excellent
operation of the KEKB accelerator. We acknowledge support from the
Ministry of Education, Culture, Sports, Science, and Technology of Japan
and the Japan Society for the Promotion of Science; the Australian
Research
Council and the Australian Department of Industry, Science and Resources; 
the Department of Science and Technology of India;
the BK21 program of the Ministry of Education of Korea and the CHEP SRC
program of the Korea Science and Engineering Foundation;
the Polish State Committee for Scientific Research under contract No.2P03B 
17017;
the Ministry of Science and Technology of Russian Federation;
the National Science Council and the Ministry of Education of Taiwan;
and the U.S. Department of Energy.

\clearpage





\begin{thebibliography}{99}

\bibitem{beneke}{M.~Beneke {\it et al.}, 
        Phys. Rev. D {\bf59}, 054003 (1999).}
%
\bibitem{diehl}{M.~Diehl and G.~Hiller, hep-ph/0105194,
         JHEP 0106:067 (2001).}
%
\bibitem{edwards}{K.W.~Edwards {\it et al.} (CLEO Collaboration),
        Phys. Rev. Lett. {\bf86}, 30 (2001).}
%
\bibitem{KEKB}{KEKB B Factory Design Report, KEK Report 95-1, 1995,
	unpublished.}
%
\bibitem{NIM}{A.~Abashian {\it et al.} (Belle Collaboration),
        KEK Report 2000-4,
	to be published in Nucl. Inst. and Meth. A.}
%
\bibitem{GEANT}{Events are generated with the CLEO group's QQ program
        (http://www.lns.cornell.edu/public/CLEO/soft/ QQ); the detector
        response is simulated with GEANT, R.Brun {\it et al.},
        GEANT 3.21, CERN Report DD/EE/84-1, 1984.}
%
\bibitem{khh}{K.~Abe~{\it et al.} (Belle Collaboration), 
        BELLE-CONF-0114; hep-ex/0107051;}
%
\bibitem{dEreg}{The negative $\Delta E$ region could contain events 
        from $B\to\chi_{c0}K\pi$, with a low momentum neutral or charged pion
        that is not included in the $M_{bc}$ and $\Delta E$ computation.
        Therefore, we do not use the $\Delta E<-0.1$~GeV region for sideband
        studies and also exclude it from the fits.}
%
\bibitem{VCal}{D.M.~Asner {\it et al.} (CLEO Collaboration),
        Phys. Rev. D {\bf 53}, 1039 (1996).}
%
\bibitem{PDG}{D.E.~Groom {\it et al.} (Particle Data Group),
        Eur. Phys. J. C {\bf15}, 1 (2000).}
%
%
\bibitem{Correct}{The quoted value for the 
        $\chi_{c0}\to K^{*0}(892)K^-\pi^+ + c.c.$ branching fraction 
        is based on the results of \mbox{MARK-I} published in:
        W.~Tanenbaum \mbox{{\it et al.}}, Phys. Rev. D {\bf17}, 1731 (1978).
        If, however, one takes into account the recent precise
        measurement of the $\chi_{c0}\to K^+K^-\pi^+\pi^-$ branching ratio
        by the BES group,
        J.Z.~Bai \mbox{{\it et al.}}, Phys. Rev. D {\bf60}, 072001 (1999), 
        the discussed branching becomes
        $(1.5 \pm 0.4)\times10^{-2}$  not affecting our conclusions.}
%
\bibitem{mark}{F.~Vannucci {\it et al.}, Phys. Rev. D {\bf15}, 1814 (1977).
        The quoted number is our estimate from the data presented in this 
        reference. From Fig.~9 therein, we estimate $120\pm13$ signal events 
        in the $K^{*0}(892)K^-\pi^+$ final state. Assuming the same 
        reconstruction efficiency as for the $K^+K^-\pi^+\pi^-$ final state,
        we obtain ${\cal{B}}(J/\psi\to K^{*0}(892)K^-\pi^+ + c.c.)=
        (6.3\pm2.1)\times10^{-3}$.}
%
\bibitem{chi0}{K.~Abe~{\it et al.} (Belle Collaboration), 
        BELLE-CONF-0138; hep-ex/0107050;}
%
\end{thebibliography}
\end{document}